\documentclass[]{spie}  

\usepackage{amsmath,amsfonts,amssymb}
\usepackage{graphicx}
\usepackage[colorlinks=true, allcolors=blue]{hyperref}

\title{Flowdown of the TMT astrometry error budget(s) to the IRIS design}

\author[a]{Matthias Sch\"ock}
\author[b]{David Andersen}
\author[a]{John Rogers}
\author[c]{Brent Ellerbroek}
\author[c]{Eric Chisholm}
\author[b]{Jennifer Dunn}
\author[b]{Glen Herriot}
\author[d]{James Larkin}
\author[e]{Anna Moore}
\author[f]{Ryuji Suzuki}
\author[e]{James Wincentsen}
\author[g]{Shelley Wright}

\affil[a]{TMT International Observatory, Victoria, BC, Canada}
\affil[b]{NRC Herzberg, Victoria, BC, Canada}
\affil[c]{TMT International Observatory, Pasadena, CA, USA}
\affil[d]{University of California, Los Angeles, CA, USA}
\affil[e]{California Institute of Technology, Pasadena, CA, USA}
\affil[f]{National Astronomical Observatory of Japan, Tokyo, Japan}
\affil[g]{University of California, San Diego, CA, USA}

\begin{document} 
\maketitle

\begin{abstract}
TMT has defined the accuracy to be achieved for both absolute and differential astrometry in its top-level requirements documents.  Because of the complexities of different types of astrometric observations, these requirements cannot be used to specify system design parameters directly.  The TMT astrometry working group therefore developed detailed astrometry error budgets for a variety of science cases.  These error budgets detail how astrometric errors propagate through the calibration, observing and data reduction processes. The budgets need to be condensed into sets of specific requirements that can be used by each subsystem team for design purposes.  We show how this flowdown from error budgets to design requirements is achieved for the case of TMT's first-light Infrared Imaging Spectrometer (IRIS) instrument.

\end{abstract}


\section{Introduction}

The Thirty Meter Telescope (TMT) has very tight requirements on the accuracy to be achieved for both differential and absolute astrometry with adaptive optics (AO) corrected images: 50 micro arcsec in a 100-second exposure with 10~$\mu$as systematic errors and 2~mas, respectively.  In order to ensure that the observatory will be able to deliver astrometric observations compliant with those requirements, we have established detailed astrometry error budgets, trying to account for all effects that might compromise the achieved accuracy.  Even these error budgets can, however, not be used directly for design purposes, as they are specified in terms of astrometric accuracy rather than the engineering parameters relevant for a given subsystem, and because of the many complex cross-connections between error terms and their causes.

We are therefore in the process of going through the error budget in the other direction, starting with the assumptions made to arrive at the error term magnitudes given in the budgets and identifying all the requirements needed to ensure that these assumptions are satisfied.  This flowdown is currently still in progress but has already significantly increased our confidence that the real TMT, when built, will permit high-accuracy astrometric observations on a fundamental level.  

\begin{figure} [t]
   \begin{center}
   \begin{tabular}{c} 
   \includegraphics[width=17cm]{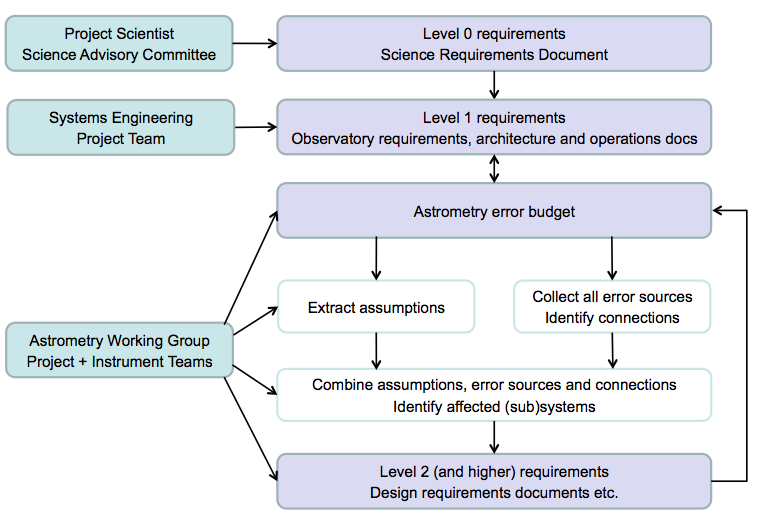}
   \end{tabular}
   \end{center}
   \caption[example] 
   { \label{f:block_diagram} 
Block diagram of the TMT astrometry error budgeting and requirements development process.  The flowdown described is this paper is what is shown from the "Astrometry error budget" box on downward.}
   \end{figure}

In this paper, we first summarize the status, content and purpose of the TMT astrometry error budgets.  Note that, unless we are referring to a specific science case, we generally use the plural form, as there is no such thing as {\it the} TMT astrometry error budget.  In Section~\ref{s:flowdown} we then describe the process by which we are verifying that requirements exist for all assumptions made in these error budgets and illustrate this with a variety of examples of different complexity (see also Fig.~\ref{f:block_diagram} for a block diagram).  We conclude by summarizing key findings to date, the current status of this work and the next steps to be taken.

\section{TMT astrometry error budgets summary}

The TMT astrometry requirements and error budgets have been described in detail previously\cite{schoeck14_astrometry_spie} and are only summarized here.  The TMT Science Requirement Document (SRD) contains two top-level requirements:

\begin{description}
\item[Differential astrometry:] The differential astrometric error shall be no more than 50~$\mu$as in a 100 second exposure in H band, and shall fall to a systematic floor of 10~$\mu$as as the integration time is increased.
\item[Absolute astrometry:] The error of absolute astrometry shall be no more than 2~mas. 
\end{description}

Other than the exposure time and bandpath given in the first requirement, the SRD does not specify for which types of astrometric observations and under what conditions these requirements must be met.  It is clear that there are situations, both environmental conditions and science field characteristics, under which this is not possible, for example, when the observed objects are too faint, when turbulence is exceptionally strong or when there are not enough reference objects in the science field to correct for low-order distortions with sufficient accuracy.  The work on the astrometry error budgets therefore not only serves the goal of ensuring that the SRD requirements can be met for the highest-accuracy science cases, but also to determine under what conditions this is possible and what accuracy can be expected for other science cases.

The TMT error budgets are thus divided into a variety of science cases.  There are three so-called generic science cases: absolute astrometry, differential astrometry of science objects relative to each other, and differential astrometry of science objects relative to a reference frame of other objects in the field.

An example of an absolute astrometry case is the measurement of the positions of jet ejecta from a star forming region in a field of view that is so obscured that no background objects can be seen.  Note that even this case is not truly absolute astrometry in the strict sense, as the ejecta positions are still calculated relative to the coordinates of the objects TMT uses to determine its pointing.  In this case, these references would be the natural guide stars (NGSs) of TMT's multi-conjugate adaptive optics (MCAO) system, NFIRAOS, used for tip/tilt and focus sensing.\footnote{We should point out at this time that we are only dealing with AO astrometry here, seeing-limited observations are not considered.}

Examples of differential astrometry science cases of objects moving relative to each other are multiple star or exoplanet systems or the motions of stars in clusters or dwarf galaxies.  These may turn into the category of differential astrometry with respect to a background reference frame when there is a sufficient number of unrelated reference objects in the frame that can be used to improve the accuracy of the measurement.

In addition, a number of specific science cases are analyzed, either because they are not described well by the generic science cases, or because they are important enough for TMT that we want to consider them individually.  One example of a specific science case is measuring the orbits of the stars moving around the super massive black hole at the Galactic center.

Each generic science case is further divided based on the number of reference objects visible in the field.  This results in not only quantitative differences, such as noise and random errors averaging out, but has fundamental consequences for the kind of data analysis that can be done as well.  For example, one of the largest fundamental error terms is due to the plate scale distortions caused by the error in our knowledge of the true and measured NGS positions in the instrument focal plane.  This term is present when there are fewer than three reference objects in the science field, but can be reduced to a usually negligible magnitude as soon as three or more reference objects are available.

\begin{figure} [t]
   \begin{center}
   \begin{tabular}{c} 
   \includegraphics[width=17cm]{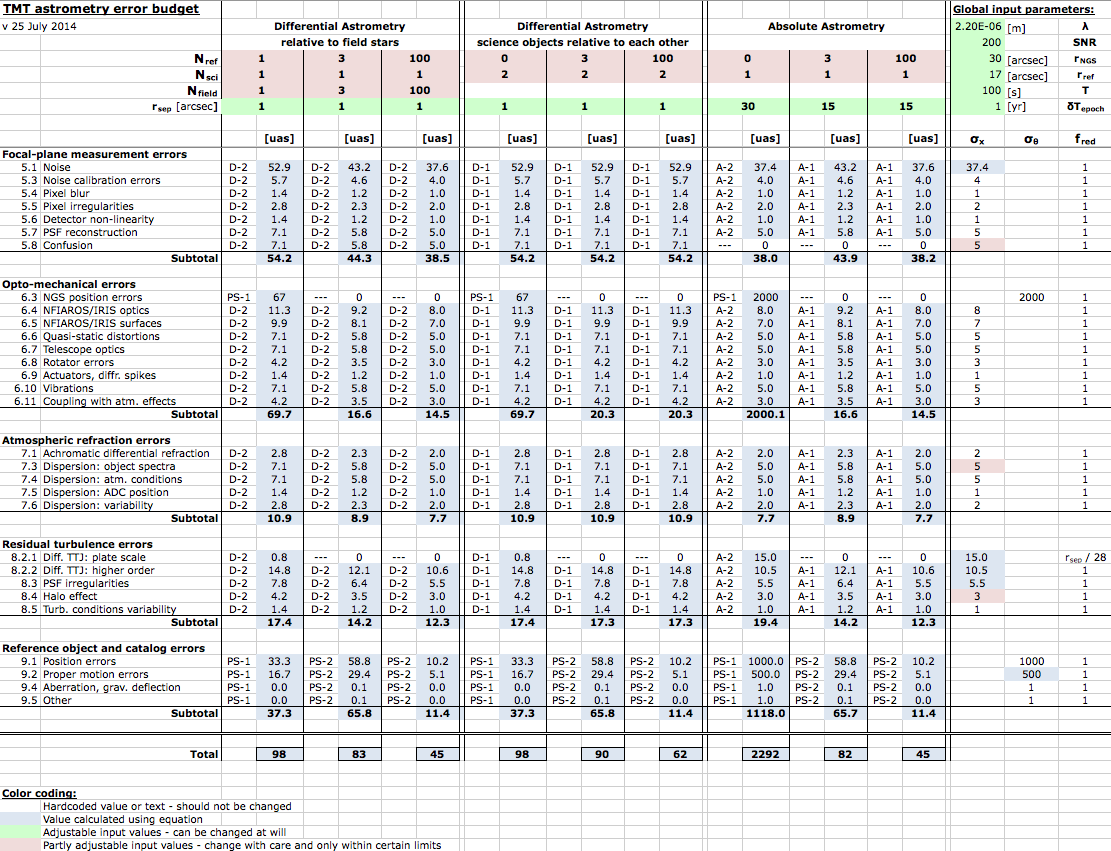}
   \end{tabular}
   \end{center}
   \caption[example] 
   { \label{f:eb} 
Example of TMT astrometry error budgets for several generic science cases.  These should be taken as examples only and not as the definitive error budgets for these science cases.  See text and references therein for details.}
   \end{figure}

Finally, for each science case, the error budget consists of a large number of error terms, usually between 30 and 40 depending on which terms are applicable.  An example of several generic science case error budgets is shown in Fig.~\ref{f:eb}.  Note that these should be taken as examples only and not as the definitive error budgets for these science cases.  Depending on the specifics of the observations, the expected accuracy might be both higher or lower than what is given in the figure even for science cases that fall into the respective categories.  Furthermore, this is still work in progress and some values have large uncertainties at this time, or are simply allocations that have not yet been backed up by analysis or design.  For example, while a vibrations error budget exists for TMT, the analysis how a given level of vibrations translates into astrometric errors has not yet been performed.

\section{Flowdown to requirements}
\label{s:flowdown}

From the beginning, the error budgets shown in Fig.~\ref{f:eb} were set up with the goal of aiding the design of the observatory and its instruments.  As such, the error terms are distributed into five categories ordered by where the errors are caused rather than by what effect they have in the data analysis process.  This has been described previously\cite{schoeck14_astrometry_spie} and we do not repeat it here.  Instead, we describe how we are using the error budgets to verify that all necessary requirements are in place and sufficiently strict to allow for the highest-accuracy astrometry science cases to be done successfully.

\subsection{Identifying the sources responsible for each error budget term}

Some of the error terms can clearly be attributed to a single source or (sub)system.  For example, the pixel blur error term in the focal-plane measurement errors category is due to charge-diffusion of the detector.  This causes a random uncertainty in determining the center of the PSF and needs to be added to other terms such as photon noise.  Thus, the error term can be uniquely attributed to the detector.  The resulting requirement, on a conceptual level, is something to the effect of: charge diffusion in the detector must be small enough that it does not cause more than a certain amount of astrometry error.  In the example of Fig.~\ref{f:eb}, this is currently set to 1~$\mu$as of measurement uncertainty for an individual object, denoted by $\sigma_x$ in the far right column of the table.  The requirement is verified through analytical calculations and simulations that translate charge diffusion values, usually given in fractions of pixels, into astrometric accuracies for a variety of observing scenarios.  We note that, for engineering purposes, the requirement for the detector should be formulated in terms of charge diffusion, not astrometric accuracy.

Even such an apparently simple and well determined error source affects other error budget terms as well.  Distortion calibration, for example, is in part done by measuring the positions of sources with known coordinates and thus establishing a distortion map.  These sources might be dense star fields (such as star clusters) or the holes in the calibration pinhole grid mask that can be inserted in the NFIRAOS entrance focal plane.  The accuracy of these calibration measurements is affected by detector charge diffusion which must therefore be taken into account for the distortion terms in the error budget (opto-mechanical errors category) as well.

This is, of course, a somewhat trivial example and is chosen mostly for illustration purposes.  As it is a random error that averages out with increasing number of photons, it affects observation and calibration procedure requirements more so than requirements on the detector itself.  However, other system characteristics might have more complex effects.  Sticking with detector properties as the next example, the next term in the error budget, pixel irregularities, can result in systematic errors for both the astrometric measurement and the distortion calibration, especially if such irregularities are correlated among close pixels.  The hope is, of course, that this will affect both the distortion calibration and the measurement analogously and can therefore be removed during data processing, but that is only true to some extent.

There are also terms in the error budget which need to be split up between different systems.  For example, the term called "NFIRAOS/IRIS surfaces" in the opto-mechanical errors category refers to the deviations of the real surfaces of the optics from the ideal (design) surfaces.  As with all distortion terms, these are measured by the distortion calibration procedures described above and taken into account during data analysis.  However, several types of residuals remain after calibration in addition to the noise/error of the calibration process itself: high-order aberrations which cannot be measured by the calibration procedure, aliasing of high-order aberrations into low-order modes and differences between the calibration and science observation configurations (for example due to beam wander on the optical surfaces).

These residuals need to be distributed between IRIS and NFIRAOS and, in fact, between the different surfaces of all the individual optics.  We have shown\cite{ellerbroek13} that this is a reasonably straight-forward process as far as specifying the surface quality requirements themselves is concerned, but only if the input assumptions are know with sufficient confidence.  For example, beam wander on the surfaces is a function of the difference in elevation and azimuth angles between the calibration setup and the science observations, of the telescope optics alignment stability and of the relative motion of the different optics in NFIRAOS and IRIS with respect to each other.  Thus, while this error term is due to the surface quality of the AO system and instrument optics, it also flows down to requirements on the mounts of the optics, the mounts of the AO system and instrument, the telescope optics and control system, and the calibration procedure.

   \begin{figure} [t]
   \begin{center}
   \begin{tabular}{c} 
   \includegraphics[width=17cm]{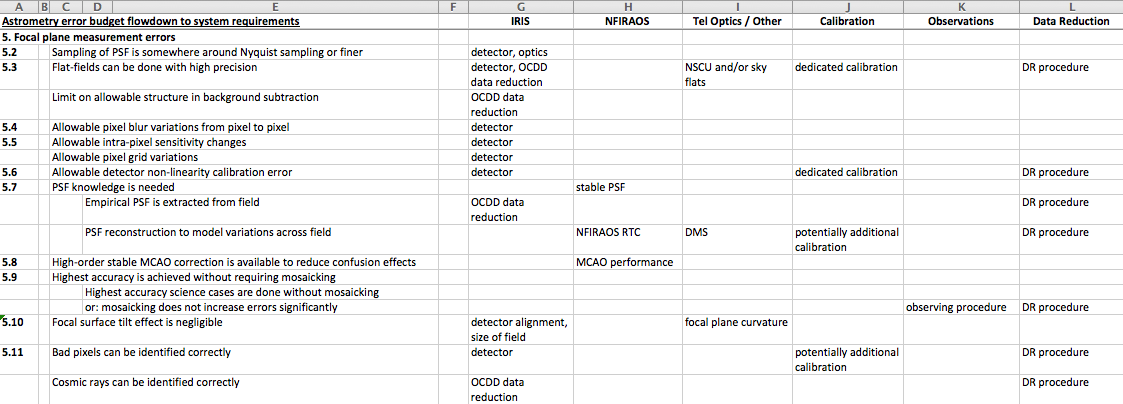}
   \end{tabular}
   \end{center}
   \caption[example] 
   { \label{f:flowdown1} 
First step of the TMT astrometry requirements flowdown process.  See text for details.  Note that this is still work in process and therefore likely incomplete.}
   \end{figure} 

   \begin{figure} [t]
   \begin{center}
   \begin{tabular}{c} 
   \includegraphics[width=17cm]{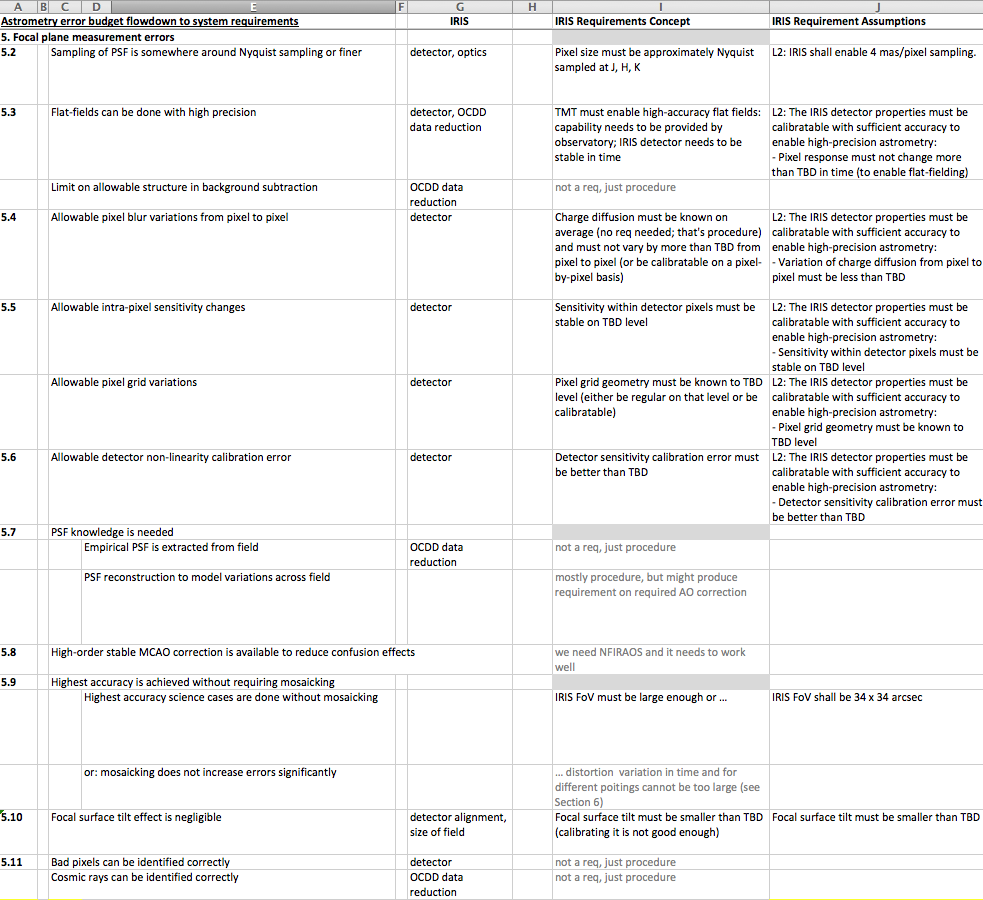}
   \end{tabular}
   \end{center}
   \caption[example] 
   { \label{f:flowdown2} 
Second step of the TMT astrometry requirements flowdown process.  See text for details.  Note that this is still work in process and therefore likely incomplete.}
   \end{figure}

\subsection{Identifying assumptions made in the error budget}

The previous sections described the error budgets and provided some examples of how error budget terms necessitate requirements even for systems that, on casual inspection, have nothing to do with that specific error.  We now describe the process by which we are attempting to capture all these connections and requirements.  We point out that, in the end, the requirements and the assumptions and analyses from which the flow down from the error budget will be captured by TMT's requirements system implemented in DOORS.  At this time we are still in the process of developing this process and most work is done in a series of informal spreadsheets.  These will then serve as the basis for the more formal systems engineering process.

In addition to being summarized in spreadsheets such as Fig.~\ref{f:eb}, the TMT astrometry error budgets are also described in much more detail in a dedicated report.  The requirement flowdown process starts by extracting all the assumptions made in this report and assigning them to the different TMT systems and sub-systems, as well as the calibration, observation and data reduction procedures that contribute to the assumptions being valid.  An example of this first step of the process is shown in Fig.~\ref{f:flowdown1}.  It captures all the assumptions made in Chapter 5, focal plane measurement errors, of the astrometry error budget report.  Note that, for this step, we do not consider whether any of these assumptions will, in the end, result in an actual requirement, and whether they are also captured somewhere else in the report.  We simply attempt to list all assumptions in the order in which they appear in the report.

Columns G to L then attribute these assumptions to all systems or procedures on which they have an effect.  The categories for this step are the instrument (IRIS), the AO system (NFIRAOS), any other hardware (grouped together into one column in this sheet for brevity, but not in follow-up steps), calibration steps needed, requirements on the science observations and on the data reduction (DR) procedure.  The content of each cell also denotes which subsystem and/or requirements document might be responsible for the resulting requirement(s).

We emphasize again that this spreadsheet, as well as those shown in the next section, are meant as working tools only.  While we obviously need them to be complete, we otherwise try to keep them as simple and informal as possible.

\subsection{Identifying system requirements}

The sheet from which Fig.~\ref{f:flowdown1} is taken contains on the order of 100 assumptions extracted from the error budget documents.  In the next step, Columns A to G of this spreadsheet are copied into a number of other sheets, one for each system or procedure type.  Roughly speaking, there is one sheet, or in some cases several, for each of Columns H to L in Fig.~\ref{f:flowdown1}.

A small section of one such sheet, the one for IRIS, is shown in Fig.~\ref{f:flowdown2}, using again the example of the focal plane errors.  The seven columns on the left are simply a copy from the previous sheet.  Column I describes in conceptual terms what this means for IRIS in terms of requirements.  Column I also contains grayed out cells and cells with text in gray font.  These are assumptions that either do not result in a requirement (for example, because they simply describe the procedure) or that apply to other systems (which are captured in other sheets).

Column J more directly states how the respective requirement {\it might} be formulated in its shortest form, without trying to use the text of actually existing requirements.  Note that, in a way, Columns I and J are redundant.  They simply demonstrate how we work our way from the assumptions on the left to increasingly more formal requirements on the right.  In fact, there is another column on the right which links these assumptions to an actual requirement, or requirements, in the IRIS design requirements document (DRD) or another applicable requirements document.  This column is not shown here because this part of the process is still on-going and has not yet been approved by the instrument team and the project, but it is well advanced and most requirements are in place.  For example, IRIS DRD Requirement 0760 requires the IRIS imager plate-scale to be within 5\% of 0.004 arcsec/pixel, corresponding to Assumption 5.2 in Fig.~\ref{f:flowdown2}.  Requirement 3130 states that the IRIS imager and integral field spectrograph detectors shall be characterized for charge diffusion, pixel sensitivity and pixel grid geometry, thus covering several of the line items in the figure.

Once the link between the assumptions on the left and official requirements has been established, the flowdown from the error budgets is complete and the results can be added to the appropriate systems engineering tools and documents.  These include the TMT-wide requirements flowdown matrix, design requirements documents for the different systems and subsystems and interface control documents.

\section{Summary}

The main purpose of creating the TMT error budgets was to establish whether, in principle, the observatory is capable of astrometric observations with the accuracies required in the science requirements document and that all contributions to the astrometric errors are accounted for.  The purpose of the requirements flowdown process described here is to ensure that, in practice, TMT as designed and built will live up to those expectations.  

The desired outcome of the process was therefore that we would discover nothing new: that all requirement assumptions shown in Column J of Fig.~\ref{f:flowdown2} have an existing counterpart in the system DRDs and that the values given in those requirements result in the observatory being compliant with the astrometry error budgets for the different science cases.  So far, no show stoppers or large discrepancies have been found.  On the other hand, the process, even at its current stage, has already assisted us in identifying some details of the design that might have otherwise gone unnoticed.  Examples are some details of the polishing specifications of the optics in NFIRAOS and IRIS, the allowable relative motion of the detector with respect to the focal plane, the distortion calibration procedure and hardware and, most recently, the design of the mounts of the NFIRAOS entrance windows.  The next steps are the completion of the flowdown process, the acceptance of the results by the different (sub)system teams and the project, and entering them into the official requirements documents and flowdown matrix.

\acknowledgments 
 
The TMT Project gratefully acknowledges the support of the TMT collaborating institutions.  They are the California Institute of Technology, the University of California, the National Astronomical Observatory of Japan, the National Astronomical Observatories of China and their consortium partners, the Department of Science and Technology of India and their supported institutes, and the National Research Council of Canada.  This work was supported as well by the Gordon and Betty Moore Foundation, the Canada Foundation for Innovation, the Ontario Ministry of Research and Innovation, the Natural Sciences and Engineering Research Council of Canada, the British Columbia Knowledge Development Fund, the Association of Canadian Universities for Research in Astronomy (ACURA) , the Association of Universities for Research in Astronomy (AURA), the U.S. National Science Foundation, the National Institutes of Natural Sciences of Japan, and the Department of Atomic Energy of India.


\end{document}